\documentstyle[prl,aps]{revtex}
\begin{document}
\title{Fracture in Three-Dimensional Fuse Networks}
\author{G.\ George Batrouni$^{1,2}$ and Alex Hansen$^3$}
\address{$^1$ Institut Non-Lin\'eaire de Nice, Universit\'e de Nice--Sophia
Antipolis, 1361 route des Lucioles, F--06560 Valbonne, France}
\address{$^2$H{\"o}chtleistungsrechenzentum, Forschungszentrum J{\"u}lich 
GMBH, D--52425 J{\"u}lich, Germany}
\address{$^3$Institutt for fysikk, Norges teknisk-naturvitenskapelige 
universitet, NTNU, N--7034 Trondheim, Norway}
\date{\today}
\maketitle
%%%%%%%%%%%%%%%%%%%%%%%%%%%%%%%%%%%%%%%%%%%%%%%%%%%%%%%%%%%%%%%%%%%%
\begin{abstract}
We report on large scale numerical simulations of
fracture surfaces using random fuse networks for two very different
disorders. There are some
properties and exponents that are different for the two distributions,
but others, notably the {\it roughness exponents,\/} seem universal. 
For the universal roughness exponent we found a value of $\zeta=0.62\pm 0.05$.
In contrast to what is observed in two dimensions, this value is lower 
than that reported in experimental studies of brittle fractures, and rules out
the minimal energy surface exponent, $0.41\pm 0.01$.
\end{abstract}
\pacs{PACS: 46.30Nz,62.20M,81.40Np}
%%%%%%%%%%%%%%%%%%%%%%%%%%%%%%%%%%%%%%%%%%%%%%%%%%%%%%%%%%%%%%%%%%%

During the past ten years or so fracture mechanics has caught the
interest of the physics community.  It is a field of
utmost technological importance, and at the same time it poses
fundamental questions on the interplay between disorder and dynamical
effects.  Early on, simple static models of brittle fracture, such as the {\it
fuse model\/} \cite{arcangelis85} were constructed and studied
extensively with the limited numerical tools of the time
\cite{herrmann90}, which meant that only small two-dimensional 
systems were accessible.  Even so, several qualitative and some
quantitative results were obtained.  Among the qualitative results, we
may quote the existence of several types of fracture regimes depending
on the type of initial disorder in the system\cite{khang88,hansen91}. The
most notable quantitative result was the roughness of cracks
obtained using the random fuse model\cite{hansen91b}.  This
roughness, defined as a typical length scale associated with the
direction perpendicular to the fracture plane, was found to scale with
the linear size of the fracture plane to a power law with a roughness
exponent $\zeta=0.70\pm 0.07$ for a wide class of different
disorders. This exponent was later measured for two-dimensional
stackings of collapsable cylinders \cite{poirier92} ($\zeta\approx
0.73$), for paper tear lines \cite{kertesz93} ($\zeta=0.68\pm0.05$)
and for fractures in thin wood plates, where the grain structure was
oriented parallel to the short axis ($\zeta=0.68\pm0.04$)\cite{engoy94}.

In three dimensions, studies of the scaling properties of fracture surfaces
started with the work of Mandelbrot et al.\ 
\cite{mandelbrot84}.  Bouchaud et al.\ \cite{bouchaud90} suggested that
the roughness exponent for brittle materials has a universal value
close to 0.8.  Later experimental work on a wide range of materials,
e.g.\ by M{\aa}l{\o}y et al.\ \cite{maloy92} gave results which were
consistent with such a hypothesis.  However, Milman et al.\
\cite{milman93} contested this hypothesis, and supported their claims
with atomic force microscope measurements of fracture surfaces in
crystalline metals.  The picture that is emerging today is one where
there is a smaller roughness exponent of about 0.4 --
0.5 at very small scales, which crosses over to a higher value
($\approx 0.8$) at larger scales \cite{bouchaud94,bouchaud95}.  The
smaller roughness exponent has been associated with slow crack
propagation where dynamical effects are negligible, while the higher
crack speeds give rise to the higher value.  It has been speculated
that the smaller value is that of the minimum-energy surfaces
\cite{bouchaud94}, which by the best estimate of today is known to be
$0.41\pm0.01$ \cite{middleton95}.  We note that in
two dimensions, the corresponding minimal-energy surface problem leads
to a roughness exponent equal to 2/3, i.e., very close to that found
in the fuse model.

Given the success of the fuse model in reproducing the experimentally
observed roughness exponents in two dimensions, it is of great
interest to test this model in three dimensions.  As the fuse model
does not contain any dynamical fracture properties, it is capable of
isolating the effect of the interplay between the stress-distribution
(modeled as an electrical current distribution) and the distribution
of local strengths in the material. It therefore models slow crack
propagation.  In particular, the hypothesis of a connection between the
roughness exponent in this regime and that of the minimal-energy surfaces 
can be tested.

The model is a three-dimensional lattice with the near neighbour
sites connected by bonds, taken to be electrical fuses with identical 
resistance but whose burn-out thresholds are disordered. A fuse conducts
until the current it carries, $i$, exceeds the burn-out (breaking) threshold, 
$i_c$, at which point it becomes irreversibly an insulator. The breaking 
thresholds, $i_c$, are drawn from some probability distribution (see below) 
and the potential difference across the $L\times L\times L$ lattice is 
applied along a diagonal of the cube, i.e.\ the $(1,1,1)$ direction. The
finite size effects associated with this choice are much smaller
than taking the potential difference along a major lattice axis. To
reduce finite size effects further, we used periodic boundary conditions
in {\it all\/} directions, i.e., we do not use actual planar  
electrodes attached to the lattice to apply the potential difference. Such 
electrodes greatly modify the behaviour of the fracture surfaces in their
neighbourhood. Instead we use ``ghost'' sites that are 
not actually on the lattice to produce the potential difference \cite{roux97}. 

So, a potential difference is applied and the weakest fuse is
broken. To recalculate the currents in the bonds after a fuse has
blown, one solves the Kirchhoff equations 
by using the conjugate gradient algorithm which
is trivial to parallelize efficiently. We did our simulations on the
Connection Machine CM5. The stopping condition was that the residual
error be less than $10^{-11}$.

The system sizes we studied went from $8^3$ to $48^3$. The number of
realizations per lattice depended on its size, on the type of
disorder used and the quantity studied. 
We list these numbers further on in the text.

The breaking thresholds were assigned by generating 
random numbers uniformly distributed on the unit interval and raising it
to a power $D$.  This corresponds to a cumulative probability distribution
$P(i_c) = {i_c}^{1/|D|}$ when $D>0$ and $P(i_c) = 1-{i_c}^{-1/|D|}$ when $D<0$. 
The smaller the value of $|D|$, the smaller the disorder.
Furthermore, when $D>0$ the distribution of strengths 
has a power law tail extending towards weak bonds (i.e.\ those with a
small threshold), while when $D<0$, the tail extends towards strong
bonds.  Depending on the value of $D$, the two-dimensional fuse model
was shown to exhibit distinct classes of fracture behavior
\cite{hansen91}.  For small values of $|D|$, a macroscopic
crack starts developing early in the fracture process, while when
$|D|$ is large, a cloud of disconnected ``microcracks" (blown fuses)
develops before they coalesce into the a final macroscopic crack.  In
the current work we study the fracture surfaces in three dimensional
systems with weak disorder, $|D|=0.5$, i.e.\ macroscopic fracture
surfaces start to develop early in the fracture process (localized
fracture). To study universality issues we examined both positive and
negative values of $D$, $D=\pm 0.5$.  The idea is that with such
widely different distributions, universal and nonuniversal properties
will be clearly distinguished.

Before turning to the question of fracture roughness, we report on some
further results concerning the fracture process.  These results should be
compared with those obtained earlier for the two-dimensional fuse model 
\cite{arcangelis89,arcangelis89b}, and for a three-dimensional elastic
bond-bending model \cite{arbabi90,sahimi93}.  
We find for $D=0.5$, that $N_b \propto
L^{2.6}$ while for $D=-0.5$ we find $N_b \propto L^{2.1}$, where $N_b$
is the total number of broken bonds. 
The exponent $2.1$ is consistent with $2$, the trivial
geometrical exponent of a surface, indicating that most of the broken
bonds for this case belong to the fracture surface. The lack of
univerality for this exponent can be understood by examining the
details of the formation of the fracture surface.  Even though both
for positive and negative $D$ the simulations are done in the
localized phases, the fracture surfaces evolve differently for each
case. For the $D> 0$, the distribution allows for quite a few very
weak bonds which must be broken before the fracture surface starts to
develop.  This initial breaking process is disorder-dominated, and
crosses over to current dominated breaking after enough bonds have
been broken. At the crossover point, the fracture surface starts to
develop and spread. But at this point there are already many broken
bonds which do not belong to the fracture surface and which are
randomly distributed in the {\it volume\/} of the lattice. On the
other hand, we found that for $D< 0$ the breaking process starts out
in the current dominated mode, for weak disorder, and so it is one of
the very first broken bonds that determine where the fracture surface
will start spreading. There are no broken bonds randomly distributed
in the volume, in marked contrast with the $D> 0$ case. This result
and its explanation are in agreement with those of
\cite{arcangelis89b} in two dimensions.

This same effect is clearly seen if one examines the evolution of
the conductivity of the networks as a function of broken bonds. Fig\
\ref{fig1} shows (for $L=8, 12, 16, 32, 48$, $D=-0.5$) the conductivity as a
function of ``time,'' $t\equiv i/L^{\omega}$, where $i$ is the number
of broken bonds. We see that the data collapse is very good for
$\omega=2.13$ throughout the fracture process (except at the very end
where the finite size effects are appreciable). On the other hand,
positive $D$ gives a very different picture. Fig.\ \ref{fig2} (insert) shows
the conductivity versus $t=i/L^3$ for the same sizes as the previous
case but for $D=0.5$. We see that for times up to $t\approx
0.25$, the data collapse is excellent. In addition we see that the
scaling exponent, $3$, simply corresponds to the geometric dimension.
As mentioned before, this is due to the fact that
for this case, the early times are disorder driven and therefore the
broken fuses are randomly distributed in the volume thus giving this
geometric exponent. However, we also see that there is a sudden
crossover to a different behaviour. This is the crossover to the
current driven regime. Fig.\ \ref{fig2} furthermore shows the development
of the conductivity but with the $t=i/L^2$ and where the curves are shifted
to the right to coincide with the current driven curve for $L=48$.  We
see that the data collapse in the current driven regime is excellent.
This shows that in the {\it current driven\/} regime the fracture
surfaces evolve at the same rate for all sizes and for both values
of $D$ (the two exponents, $2$ and $2.13$ are consistent). 
For $D>0$ we find, in addition, that the
number of bonds to be broken in order to cross over to the
current dominated regime scales like $L^{2.7}$. The value found
in~\cite{arcangelis89b} is $1.65\pm 0.03$ for the two dimensional
case. This is consistent with our result in one additional dimension.

Fig.\ \ref{fig3} shows the $I-V$ curves for $D=\pm 0.5$ and $L=8, 12,
16, 32,$ and $48$. It shows clearly that the data follow the scaling
law \cite{arcangelis89} $I=L^{\alpha}f(VL^{-\beta})$,
where we found the best data collapse for $\alpha=2$ and $\beta=1$.
These values are close to those found in \cite{arcangelis89b} for two
dimensional systems, except that one must add $1$ to $\alpha$.  This,
of course, is not surprising since in three dimensions the current
passes through a surface rather than a line as in two dimensions. The
universality of these exponents is very clear since the collapse is
excellent for both $D=0.5$ and $D=-0.5$. In addition, Fig.\ \ref{fig3}
shows that the two $I-V$ characteristics are complimentary in the
sense that the slopes of the scaling parts are identical. 

Finally we get to the roughness of the fracture surfaces. Our
three-dimensional lattice has the topology of an $S^3$ torus.  We cut
it open in such a way that the fracture surface forms a square sheet
and orient the surface in such a way that the $z$-axis points along
the $(1,1,1)$ direction, i.e., the mean current direction before any
fuse has burned out.  We measure the typical length scale $W$ in the
$z$-direction using three different norms, (1) $||\Delta
z||_2=\sqrt{(\sum_i z_i^2)/N_s-(\sum_i z_i/N_s)^2}$, where $z_i$ is
the $z$-coordinate of the $i$th broken bond belonging to the fracture
surface and $N_s$ is the total number of these, (2) $||\Delta
z||_{\infty}=(\max_i z_i-\min_i z_i)$, and (3) the smallest eigenvalue
of the moment of inertia tensor of the fracture surface, $I_{\min}$.
Fig.\ \ref{fig4} shows $W$ as a function of $L$ based on these three
norms for the two types of disorder. For $D=0.5$, lattice sizes are
$L=8, 12, 16, 32, 48$ with $600, 200, 402, 146, 133, 37$ realizations
respectively, and for $D=-0.5$, the sizes are $L=8, 12, 16, 32, 48$
with $300, 200, 200, 64, 21$ realizations respectively. The figure
shows that there is no appreciable difference in roughness between the
two types of disorder used.  There are strong finite-size corrections
to the power laws that the different kinds of roughness measures
follow.  However, the way these corrections affect the asymptotic
power laws depends on the measures.  For the $||\Delta z||_\infty$ the
effective roughness exponent approaches the asymptotic one from above,
while for the $||\Delta z||_2$ and $I_{\min}$ measures, the asymptotic
roughness exponent is approached from below.  In Fig.\ \ref{fig5}, we
show the effective roughness exponents, defined as $[\ln W(L')-\ln
W(L)]/[\ln L' -\ln L]$, based on these measures for both the $D=0.5$
and $D=-0.5$ disorders, plotted against $L'^{-0.75}$.  Two straight
lines have been added, which somewhat follow the $||\Delta
z||_\infty$, $D=0.5$ data and the $||\Delta z||_2$, $D=0.5$ data.
They act as guides to the eye.  Based on this figure, we suggest an
asymptotic rougness exponent equal to $\zeta=0.62\pm0.05$ for both
disorders.

We see that this value is smaller than the value of 0.8 typically seen
in large scale brittle fractures.  It is, however, also larger than
the value seen at small scales, which is of the order 0.4 -- 0.5.  Our result,
$\zeta=0.62\pm0.05$, seems to rule out the minimal energy exponent, 
$0.41\pm0.01$, advocated for slowly propagating cracks.  Thus, the coincidence
between the roughness exponent found in the two-dimensional
fuse model and the corresponding minimal energy problem seems
fortuitous.

We thank S.\ Roux for many valuable discussions. We also thank Thinking 
Machines Corp.\ for (almost unlimited) computer time on their CM5s. 
Travel support was provided by the CNRS and the NFR through a {\it PICS\/}
grant.
%%%%%%%%%%%%%%%%%%%%%%%%%%%%%%%%%%%%%%%%%%%%%%%%%%%%%%%%%%%%%%%%

%%%%%%%%%%%%%%%%%%%%%%%%%%%%%%%%%%%%%%%%%%%%%%%%%%%%%%%%%%
\begin{figure}
\caption{\label{fig1} Conductivity versus $t=i/L^{2.13}$ for $L=8$ (200 
realizations), 12 (200 realizations), 16 (200 realizations), 32 
(64 realizations), and 48 (21 realizations). Here $D=-0.5$.}
\end{figure}
\begin{figure}
\caption{\label{fig2} 
Conductivity versus $t=i/L^2$ plus a shift making all curves collapse
onto the $L=48$ data late in the fracture process.  The insert shows the
conductivity as a function of $t=i/L^3$.  The figures are based on 
$L=8$ (200 realizations), 12 (200 realizations), 16 (144 realizations), 32 
(62 realizations), and 48 (20 realizations). Here $D=0.5$.}
\end{figure}
\begin{figure}
\caption{\label{fig3} The I-V characteristic for $D=0.5$ and $D=-0.5$. The 
vertical axis shows $I/L^2$ as a function of $V/L$ for $L=8$, 12, 16, 32, 
and 48. Excellent data collapse is seen for the increasing part of the data. 
The number of configurations is the same as Figs.\ 1 and 2.}
\end{figure}
\begin{figure}
\caption{\label{fig4} Roughness $W$ as a function of linear lattice size $L$,
where $W$ has been estimated from (1) $||\Delta z||_2$, $D=0.5$ (circles)
and $D=-0.5$ (plus), (2) $||\Delta z||_\infty$, $D=0.5$ (crosses) and
$D=-0.5$ (squares) and (3) $I_{\min}$, $D=0.5$ (triangles) and $D=-0.5$
(diamonds). The slope of the two straight lines is 0.62.}
\end{figure}
\begin{figure}
\caption{\label{fig5} Effective roughness exponent, $\zeta_{eff}$ as a 
function of $L$ based on (1) $||\Delta z||_2$, 
$D=0.5$ (circles) and $D=-0.5$ (plus), (2) $||\Delta z||_\infty$, $D=0.5$
(squares) and $D=-0.5$ (crosses), and (3) $I_{\min}$, $D=0.5$ (triangles) and
$D=-0.5$ (diamonds). The two straight lines are guides to the eye.}
\end{figure}
%%%%%%%%%%%%%%%%%%%%%%%%%%%%%%%%%%%%%%%%%%%%%%%%%%%%%%%%%%%%%%%%%%%
\end{document}